\documentclass[aps,prl,superscriptaddress,twocolumn,reprint,amsmath,amssymb]{revtex4-2}
\usepackage[english]{babel}
\usepackage{graphicx}
\usepackage[dvipsnames]{xcolor}
\usepackage[colorlinks=true,urlcolor=blue,linkcolor=blue,citecolor=blue]{hyperref}
\usepackage{epstopdf}
\usepackage{dcolumn}
\usepackage{bm}
\usepackage{epsfig}
\usepackage{soul}
\usepackage{multirow}
\usepackage{gensymb}
\def \sample {[Cu(pz)$_2$(2-HOpy)$_2$](PF$_6$)$_2$}

\begin{document}

\title{Field-tunable Berezinskii-Kosterlitz-Thouless correlations in a 
Heisenberg magnet}

\author{D.~Opherden}
\affiliation{Hochfeld-Magnetlabor Dresden (HLD-EMFL) and W\"{u}rzburg-Dresden Cluster of Excellence ct.qmat, Helmholtz-Zentrum Dresden-Rossendorf, 01328 Dresden, Germany}

\author{M.~S.~J.~Tepaske}
\affiliation{Physikalisches Institut, Universit\"{a}t Bonn, Nussallee 12, 53115 Bonn, Germany}
\affiliation{Max Planck Institute for the Physics of Complex Systems, 01187 Dresden, Germany}

\author{F.~B\"{a}rtl}
\affiliation{Hochfeld-Magnetlabor Dresden (HLD-EMFL) and W\"{u}rzburg-Dresden Cluster of Excellence ct.qmat, Helmholtz-Zentrum Dresden-Rossendorf, 01328 Dresden, Germany}
\affiliation{Institut f\"{u}r Festk\"{o}rper- und Materialphysik, TU Dresden, 01062 Dresden, Germany}

\author{M.~Weber}
\affiliation{Max Planck Institute for the Physics of Complex Systems, 01187 Dresden, Germany}

\author{M.~M.~Turnbull}
\affiliation{Carlson School of Chemistry, Clark University, Worcester, MA 01610, USA}

\author{T.~Lancaster}
\affiliation{Durham University, Centre for Materials Physics, South Road, Durham DH1 3LE, UK}

\author{S.~J.~Blundell}
\affiliation{Clarendon Laboratory, Department of Physics, University of Oxford, Park Road, Oxford OX1 3PU, UK}

\author{M.~Baenitz}
\affiliation{Max Planck Institute for Chemical Physics of Solids, 01187 Dresden, Germany}

\author{J.~Wosnitza}
\affiliation{Hochfeld-Magnetlabor Dresden (HLD-EMFL) and W\"{u}rzburg-Dresden Cluster of Excellence ct.qmat, Helmholtz-Zentrum Dresden-Rossendorf, 01328 Dresden, Germany}
\affiliation{Institut f\"{u}r Festk\"{o}rper- und Materialphysik, TU Dresden, 01062 Dresden, Germany}

\author{C.~P.~Landee}
\affiliation{Department of Physics, Clark University, Worcester, MA 01610, USA}

\author{R.~Moessner}
\affiliation{Max Planck Institute for the Physics of Complex Systems, 01187 Dresden, Germany}

\author{D.~J.~Luitz}
\affiliation{Physikalisches Institut, Universit\"{a}t Bonn, Nussallee 12, 53115 Bonn, Germany}
\affiliation{Max Planck Institute for the Physics of Complex Systems, 01187 Dresden, Germany}

\author{H.~K\"{u}hne}
\email[Corresponding author. E-mail: ]{h.kuehne@hzdr.de}
\affiliation{Hochfeld-Magnetlabor Dresden (HLD-EMFL) and W\"{u}rzburg-Dresden Cluster of Excellence ct.qmat, Helmholtz-Zentrum Dresden-Rossendorf, 01328 Dresden, Germany}

\date{\today}

\begin{abstract}

We report the manifestation of field-induced Berezinskii-Kosterlitz-Thouless (BKT) correlations in the weakly coupled spin-1/2 Heisenberg layers of the molecular-based bulk material \sample. Due to the moderate intralayer exchange coupling of $J/k_\mathrm{B} = 6.8$~K, the application of laboratory magnetic fields induces a substantial $XY$ anisotropy of the spin correlations. 
Crucially, this provides a significant BKT regime, as the tiny interlayer exchange $J^\prime / k_\mathrm{B} \approx 1$~mK only induces 3D correlations upon close approach to the BKT transition with its exponential growth in the spin-correlation length. 
We employ nuclear magnetic resonance and $\mu^{+}$SR measurements to probe the spin correlations that determine the critical temperatures of the BKT transition as well as that of the onset of long-range order. Further, we perform stochastic series expansion quantum Monte Carlo simulations based on the experimentally determined model parameters. Finite-size scaling of the in-plane spin stiffness yields excellent agreement of critical temperatures between theory and experiment, providing clear evidence that the nonmonotonic magnetic phase diagram of \sample\ is determined by the field-tuned $XY$ anisotropy and the concomitant BKT physics.
\end{abstract}
\maketitle

Cooperative behavior and critical phenomena of strongly correlated magnets are typically dictated by the lattice and spin dimensions, as well as by the symmetry of the underlying Hamiltonian \cite{onsager_crystal_1944, mermin_absence_1966, jongh_ed._magnetic_1990, widom_equation_1965, fisher_renormalization_1974, wilson_renormalization_1975, feigenbaum_universal_1983, anderson_resonating_1987}. 
Among the most fascinating examples are two-dimensional (2D) $XY$ spin systems, which are known to undergo a topological Berezinskii-Kosterlitz-Thouless phase transition at a finite temperature $T_\mathrm{BKT}$ \cite{berezinskii_destruction_1971, kosterlitz_ordering_1973, kosterlitz_ordering_1973, kosterlitz_critical_1974},
which marks the binding of topological defects in vortex-antivortex pairs.
So far,
experimental efforts to probe a genuine BKT transition in a bulk material were compromised by the onset of 3D order 
\cite{liu_low-temperature_1990, majlis_dimensional_1992, majlis_dimensional_1993, siurakshina_theory_2000, ding_could_1992, ding_antiferromagnetic_1990, cuccoli_detection_2003} due to the inherent 3D nature of these materials.
Still, if the perturbative terms relative to a purely 2D $XY$ model are small enough, the experimental observation of magnetic properties associated with BKT correlations may be possible in the transition regime \cite{suh_spin_1995, waibel_determining_2015, klyushina_signatures_2021, Caci2021}. 

In particular, a controlled tuning of the $XY$ anisotropy, with associated impact on $T_\mathrm{BKT}$, can provide an ideal test bed for experimental studies of BKT physics and their comparison to numerical state-of-the-art modeling.
As a possible approach to tune the magnetic correlations away from 2D Heisenberg to a 2D $XY$ symmetry, the application of a uniform magnetic field to the 2D quantum Heisenberg antiferromagnet breaks the $O(3)$ symmetry, but preserves the easy-plane $O(2)$ symmetry, as was confirmed by quantum Monte Carlo (QMC) calculations \cite{cuccoli_field-induced_2003}.
Correspondingly, for Zeeman energies of the order of the exchange energy, the effective $XY$-exchange anisotropy can be controlled.
The associated BKT transition persists for all fields below saturation, yielding a nonmonotonic magnetic phase diagram \cite{cuccoli_field-induced_2003}.

\begin{figure}[!ht]
	\includegraphics[width=0.95\linewidth]{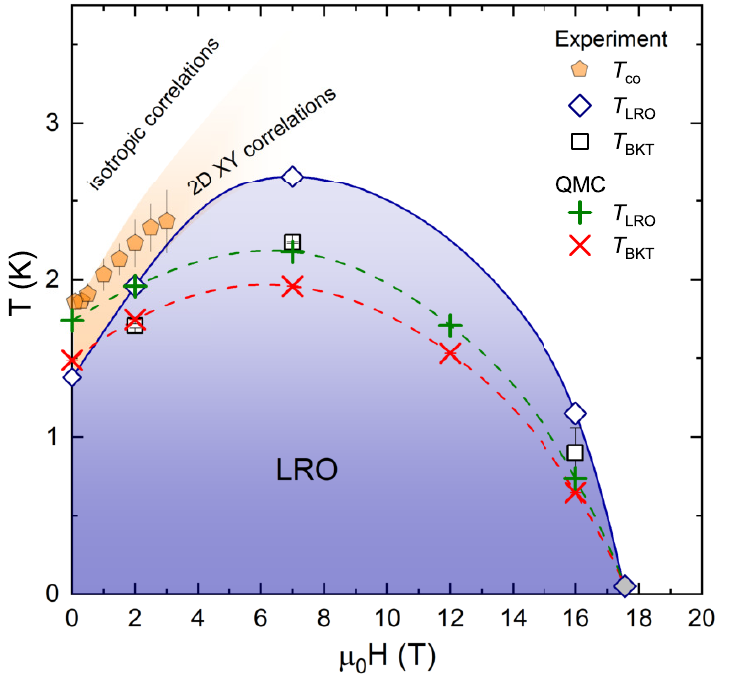}
	\caption{Phase diagram of CuPOF for out-of-plane magnetic fields from experiment and numerics. The pentagons denote the spin-anisotropy crossover temperature $T_\mathrm{co}$ from Ref.~\cite{opherden_extremely_2020}. 
White diamonds indicate the transition temperature $T_\mathrm{LRO}$ to long-range order, and squares show 
 the BKT transition temperature $T_\mathrm{BKT}$, as obtained from the analysis of the $^{31}$P $1/T_1$ rate (Fig.~\ref{fig:31P_1/T1}). $T_\mathrm{LRO}$ at zero field is determined by $\mu^{+}$SR measurements \cite{opherden_extremely_2020}. The green pluses and red crosses denote $T_\mathrm{LRO}$ and $T_\mathrm{BKT}$, respectively, as obtained from QMC calculations (Fig.~\ref{fig:QMC_FSS}). The diamond at $17.5\,\mathrm{T}$ denotes the saturation field, which was determined from magnetization experiments \cite{opherden_extremely_2020}, and is in agreement with QMC results. All lines are guides to the eye.}
	\label{fig:phase_diagram}
\end{figure}

In order to find materials which allow to study this phenomenology, the chemical engineering of molecular-based bulk magnets is a promising approach. By an appropriate choice of molecular ligands and counterions, the syntheses of several materials that realize a 2D spin-$1/2$ Heisenberg model on the square lattice were reported \cite{woodward_two-dimensional_2007, goddard_experimentally_2008, xiao_two-dimensional_2009, manson_characterization_2009, cizmar_magnetic_2010, steele_magnetic_2011, kohama_field-induced_2011, goddard_dimensionality_2012, goddard_control_2016, selmani_extremely_2010, opherden_extremely_2020}.
In these materials, a moderate nearest-neighbor exchange interaction of the order of a few K allows for the tunability of the effective exchange anisotropy by experimentally accessible magnetic fields.
Indeed, for several Cu$^{2+}$-based molecular materials, a nonmonotonic magnetic phase diagram as a function of the external field was reported \cite{cizmar_magnetic_2010, kohama_field-induced_2011, sengupta_nonmonotonic_2009, fortune_magnetic-field-induced_2014, opherden_extremely_2020}.
The magnetic properties of these molecular-based 2D quantum Heisenberg antiferromagnets were mostly investigated by thermodynamic methods \cite{woodward_two-dimensional_2007, cizmar_magnetic_2010, kohama_field-induced_2011, goddard_control_2016, selmani_extremely_2010}, thus missing local information about the magnetic correlations in the BKT transition regime.

In this Letter, we report on the field-tunable anisotropy of magnetic correlations in {\sample} [with pz = C$_4$H$_4$N$_2$, 2-HOpy = C$_5$H$_4$NHO] (CuPOF in the following), ranging from the almost-isotropic Heisenberg limit at zero field to a substantial $XY$ anisotropy upon increasing the magnetic field strength. We use nuclear magnetic resonance (NMR), as well as muon spin relaxation ($\mu^{+}$SR) as experimental probes for the dynamic and quasi-static spin correlations. Furthermore, by QMC simulations, we calculate the in-plane spin stiffness, which we use to determine the critical temperatures of the long-range order (LRO) and the BKT transition. 
Our main findings are (i) that the temperature dependence of the nuclear spin-lattice relaxation rate follows the behavior predicted from 2D BKT theory in a wide range of temperatures, determined by the field-driven anisotropy, (ii) that finite-size scaling of the QMC results permits the extraction of $T_{\text{BKT}}$, which lies below the actual 3D ordering temperature $T_{\text{LRO}}$, and (iii) that both temperatures exhibit a nonmonotonic field dependence, which is analogous to the behavior when instead of the field the anisotropy of interactions is tuned, a clear signature for the tunability of BKT correlations.

\begin{figure*}[!ht]
	\includegraphics[width=\linewidth]{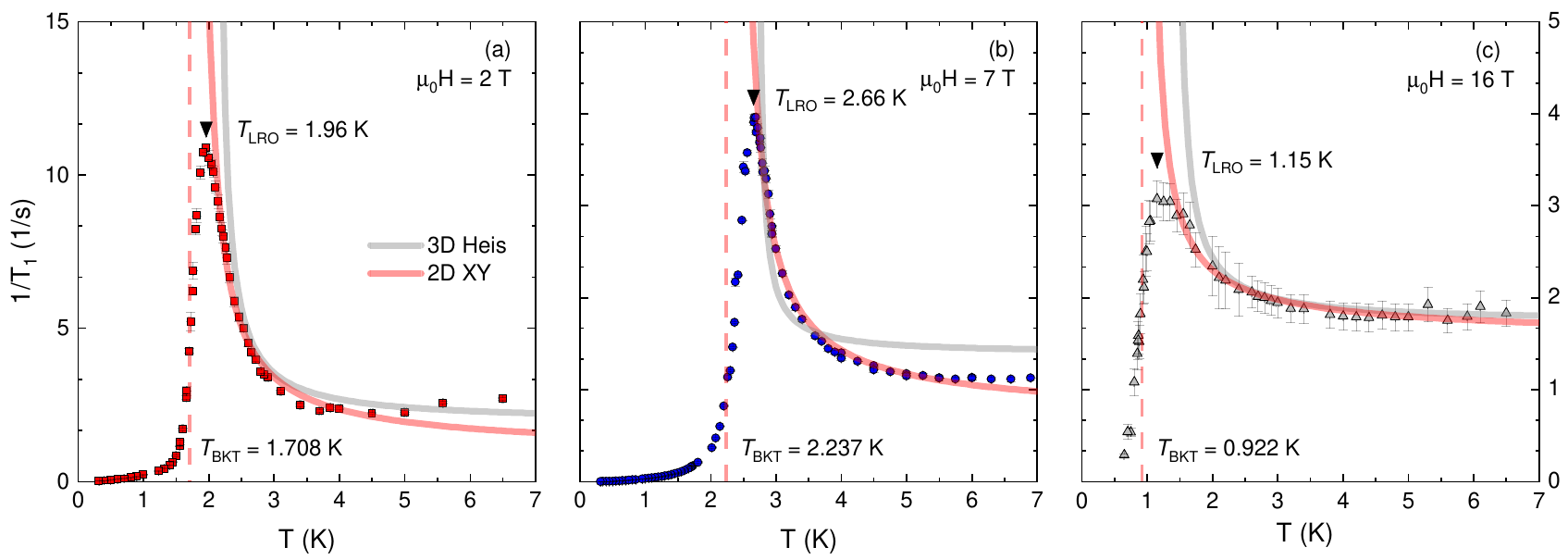}
	\caption{(a)--(c) Temperature-dependent $^{31}$P nuclear spin-lattice relaxation rate $1/T_{1}$ of CuPOF, recorded at out-of-plane fields of 2, 7, and 16~T. The solid lines are best fits according to $1/T_1\propto\xi^{z-\eta}$ for the temperature dependent correlation lengths $\xi_\mathrm{3DHeis}$ and $\xi_\mathrm{2DXY}$ of the 3D Heisenberg and the 2D $XY$ cases (see main text). The transition temperature $T_\text{LRO}$, marked with a downward triangle, is inferred from the $1/T_1$ peak position, and $T_\text{BKT}$, marked with a dotted line, is determined from fits according to $1/T_1\propto\xi_\mathrm{2D XY}^{z-\eta}$. At all fields, but most noticeably at 7 T, $1/T_1$ is described best by $\xi_\mathrm{2D XY}$ at $T \agt T_\mathrm{LRO}$.}
	\label{fig:31P_1/T1}
\end{figure*}

The synthesis and characterization of CuPOF by means of various techniques, including the $\mu^{+}$SR experiments, are described in Ref.~\cite{opherden_extremely_2020}.
The NMR spectra and spin-lattice relaxation time $T_1$ were recorded using a standard Hahn spin-echo pulse sequence and an inversion-recovery method, respectively.
The measurements were performed using a commercial phase-coherent spectrometer and a 16~T superconducting magnet, equipped with a $^3$He sample-in-liquid cryostat.

The magnetic interactions of CuPOF in an applied field are well approximated by the effective Hamiltonian
\begin{multline}
\mathcal{H}=
J\sum_{\left\langle i,j \right\rangle_\parallel }\left[S_i^xS_j^x+S_i^yS_j^y+\left(1-\Delta\right)S_i^zS_j^z\right]\\
+J^\prime\sum_{\left\langle i,j \right\rangle_\perp}\mathbf{S}_i \cdot \mathbf{S}_{j}-g\mu_B\mu_0H\sum_i S_i^z \, , 
\label{eq:hamiltonian}
\end{multline}
where $\left\langle i,j \right\rangle_\parallel$ and $\left\langle i,j \right\rangle_\perp$ denote the intra- and interlayer nearest-neighbors, and $J$ and $J^\prime$ are the intra- and interlayer exchange couplings, estimated as $J/k_\mathrm{B} = 6.8$~K and $J^\prime / k_\mathrm{B} \approx 1$~mK \cite{opherden_extremely_2020}. Whereas $\Delta = 0$ corresponds to the isotropic Heisenberg case, $0<\Delta \leq 1$ quantifies an easy-plane anisotropy, with a zero-field value of $\Delta \approx 0.01\dots 0.02$ for CuPOF \cite{opherden_extremely_2020}.

In the presence of interlayer interactions, any non-frustrated magnetic quasi-2D lattice inevitably undergoes a transition to long-range order at low temperatures.
Due to the very large separation of the magnetic layers in CuPOF, with $J^\prime/J \approx 1.4 \cdot 10^{-4}$, the very small entropy change associated with the transition to LRO is beyond the experimental resolution of thermodynamic quantities \cite{opherden_extremely_2020, sengupta_specific_2003}.
On the other hand, $\mu^+$SR is very sensitive to the local staggered magnetization, and was used to probe the transition to LRO at 1.38(2)~K in CuPOF \cite{opherden_extremely_2020}.
This transition occurs under the influence of the weak intrinsic easy-plane anisotropy, which yields a temperature-driven crossover from isotropic to $XY$-type correlations at the crossover temperature $T_\mathrm{co} > T_\mathrm{LRO}$.
An applied magnetic field increases the effective $XY$ anisotropy, which manifests itself as a field-dependent minimum of the uniform bulk susceptibility at $T_\mathrm{co}$, as depicted by the pentagons in Fig.~\ref{fig:phase_diagram}.

The temperature dependence of the $^{31}$P-NMR spin-lattice relaxation rate at out-of-plane fields up to 16~T is presented in Figs.~\ref{fig:31P_1/T1}(a)--(c).
The spin-lattice relaxation rate $1/T_1$ 
has sharp maxima at $T_\mathrm{LRO} = 1.96$ and 2.66~K at 2 and 7~T, respectively.
In comparison, the maximum amplitude of $1/T_1$ at 16~T ($T_\mathrm{LRO} = 1.15$~K) is substantially reduced. 
The transition temperatures between the 2D $XY$ and the LRO regimes are depicted by diamonds in Fig.~\ref{fig:phase_diagram}.
The strong dependence of $T_\mathrm{LRO}$ on the field strength that we observe in CuPOF clearly indicates a field-tunability of the $XY$ anisotropy of the spin correlations \cite{cuccoli_field-induced_2003}. 
This behavior is confirmed by our QMC simulations.

As previously reported, the $^{31}$P $1/T_1$ rate in CuPOF yields several broad maxima at high temperatures, which are associated with a freezing of the PF$_6$ molecular reorientation modes \cite{opherden_freezing_2021}. 
Below about 10~K, in the range of interest in the present study, these modes are frozen out and $1/T_1$ becomes temperature independent, indicating predominantly paramagnetic fluctuations.
In 2D magnetic lattices, the onset of short-range spin correlations occurs at temperatures $T \simeq J/k_\mathrm{B}$ \cite{sengupta_specific_2003}, with a correlation length of about one magnetic-lattice constant \cite{ding_spin_1990, ding_could_1992}.

At temperatures above the onset of LRO, $1/T_{1}$ can serve as a probe for the dynamic correlation length $\xi$ \cite{borsa_$35mathrmcl$_1992, suh_spin_1995, tabak_35cl_1993, bossoni_nmr_2011, gaveau_magneticfield_1991, waibel_determining_2015}. As was shown from dynamical scaling arguments \cite{borsa_$35mathrmcl$_1992}, $1/T_{1}$ is proportional to the transverse spin correlation length as $1/T_{1} \propto \xi ^{z-\eta}$, where $z$ and $\eta$ are characteristic dynamic and critical exponents \cite{jongh_ed._magnetic_1990, borsa_$35mathrmcl$_1992, suh_spin_1995, hohenberg_theory_1977}. By comparing the temperature dependence of $1/T_{1}$ with the characteristic $\xi$  of different universality  classes, we can therefore probe the nature of the predominant correlations in the critical regime, before the system finally undergoes the transition to long-range order. 
Thus, we compare the BKT correlation length of a 2D easy-plane  antiferromagnet, $\xi_\mathrm{2DXY} \propto \exp \big( 0.5 \pi/\sqrt{ T/T_\mathrm{BKT}-1} \big)$ \cite{ding_could_1992, kosterlitz_critical_1974}, with that of a 3D isotropic Heisenberg antiferromagnet, $\xi_\mathrm{3DHeis} \propto |T-T_\mathrm{LRO}|^{-0.7112}$ \cite{campostrini_critical_2002, sandvik_critical_1998}.

To describe $1/T_{1}$ in the interval $T_\mathrm{LRO} \leq T \leq J/k_\mathrm{B}$, we note that $\eta=0.0375$ for the  3D Heisenberg antiferromagnet \cite{campostrini_critical_2002}, with the LRO transition residing in the $O(3)$ universality class, whereas the easy-plane model has $\eta = 1/4$ \cite{nelson_universal_1977, makivic_low_1992, ding_phase_1992}. For both models, we use $z=d/2$, with $d$ the spatial dimensionality \cite{hohenberg_theory_1977}. The experimental estimates of $T_\text{BKT}$ are obtained from fits to the 2D $XY$ form. 

In Figs.~\ref{fig:31P_1/T1}(a)--(c), we show the measured $1/T_1$ along with both fits, for fields of 2, 7, and 16 T \cite{footPM}. In contrast to the 3D Heisenberg description, the 2D $XY$ fit accurately captures the increase of $1/T_1$ near $T_\text{LRO}$, most noticeably at 7 T. The fits yield $T_\mathrm{BKT} =$ 1.708(14), 2.237(7), and 0.90(16)~K for applied fields of 2, 7, and 16~T, respectively.
The nonmonotonic dependence of $T_\text{BKT}$ on the  field tracks that of $T_\text{LRO}$, being separated by a few hundred mK for the most part, as shown in the phase diagram in Fig.~\ref{fig:phase_diagram}. One should note, however, that the BKT transition is preempted by the LRO that arises from the 3D correlations, stemming from the finite interlayer exchange interaction $J^\prime$. In the supplemental material (SM) \cite{supplemental}, we discuss indications that changing the field strength has similar effects on the spin correlations as changing the exchange anisotropy $\Delta$ \cite{ding_could_1992, cuccoli_quantum_2003} and argue that hence the field allows to tune the effective anisotropy.

\begin{figure}[tbp]
	\includegraphics[width=0.90\linewidth]{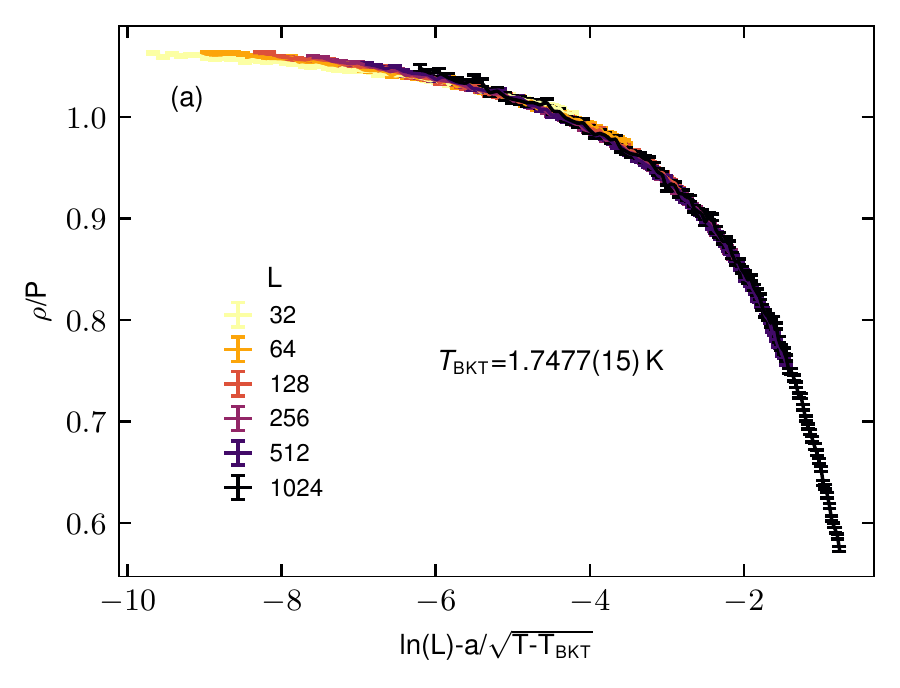}
	\includegraphics[width=0.90\linewidth]{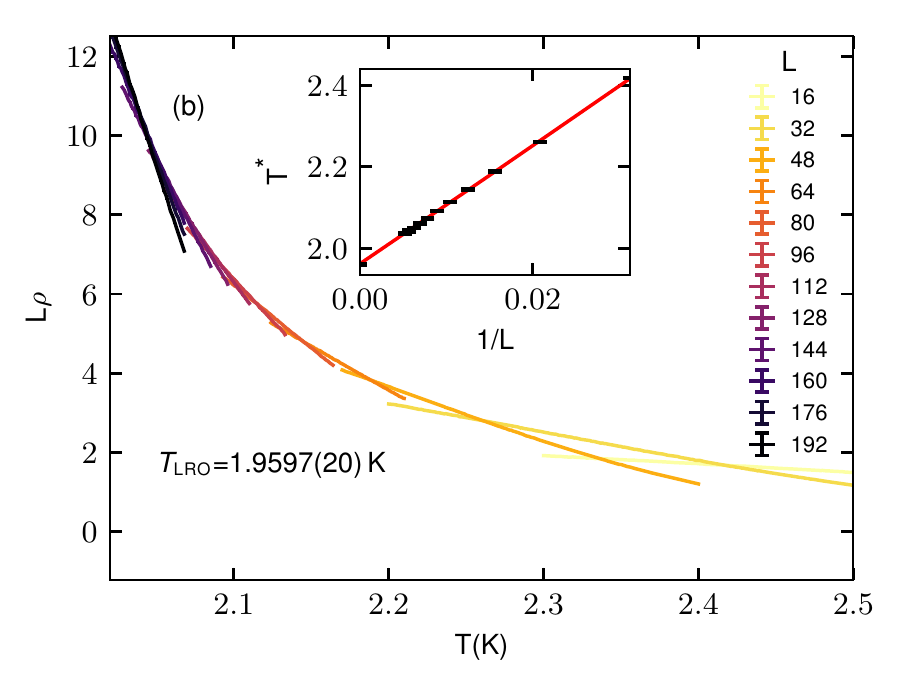}
	\caption{Finite-size scaling analysis performed to obtain the critical temperatures $T_\mathrm{BKT}$ and $T_\mathrm{LRO}$ from the QMC simulations at 2 T. (a) Data collapse of the finite-size in-plane spin stiffness $\rho$ fit closely above $T_\mathrm{BKT}$, for the $J'=0$ model, which should collapse to a single curve if the fit is perfect, reaffirming the calculated $T_\mathrm{BKT}$. The different curves correspond to different linear sizes $L$. (b) Crossings of the $L\rho$ curves for the $J'=1$~mK model; the inset shows the $L\rightarrow\infty$ scaling of the crossing temperature $T^*$. The red line denotes a second-order polynomial fit, which is extrapolated to $1/L\to0$ to estimate $T_\mathrm{LRO}$.}
	\label{fig:QMC_FSS}
\end{figure}

In order to shed more light on the experimentally observed phenomenology of mixed N\'eel and BKT-type correlations, we numerically investigate the Hamiltonian (\ref{eq:hamiltonian}) using stochastic series expansion quantum Monte Carlo with directed loops \cite{sandvik_quantum_2002}. We consider finite simple-cubic lattices with periodic boundary conditions and dimensions $L\times L\times L/8$, fixing $J/k_\mathrm{B} = 6.8$~K, $J^\prime / k_\mathrm{B} = 1$~mK, and $\Delta = 0.0185$. 
To determine $T_\mathrm{BKT}$ and $T_\mathrm{LRO}$, we calculate the in-plane spin stiffness $\rho = \left.8 L^{-3}\partial^2F/\partial\phi^2\right|_{\phi=0}$,
which is defined as the second derivative of the free energy $F$ with respect to a uniform in-plane twist angle $\phi$ \cite{sandvik_computational_2010, hsieh_finite_2013}. This quantity is non-zero in the BKT phase and in the thermodynamic limit it should vanish instantly at $T_\mathrm{BKT}$. For the finite lattices simulated with QMC, this drop-off is instead continuous, but based on how $\rho$ approaches the instant drop-off with increasing system size, we can determine $T_\mathrm{BKT}$. In particular, using finite-size scaling theory, it is predicted that $\rho$ depends on temperature $T$ and system size $L$ as \cite{hsieh_finite_2013}
\begin{align}
 \rho(T,L)/P(L) = f\left(\ln(L)-\frac{a}{\sqrt{T-T_\mathrm{BKT}}}\right), \\
 P(L) = 1+\frac{1}{2\ln(L)+c+\ln[c/2 + \ln L]} + \frac{b}{\ln^2 L} \, ,
 \label{eq:QMC_spin_stiffness}
\end{align}
where $a,b,c$ are fitting constants and $f$ is a general continuous function which we choose to be a fifth-order polynomial. This parameterization of $\rho$ is fitted closely above $T_\mathrm{BKT}$ for simulation data of the $J'=0$ model to deduce $T_\mathrm{BKT}$. Afterwards, we plot $\rho/P$ versus $\ln(L)-a/\sqrt{T-T_\mathrm{BKT}}$ in the fitting interval, which should collapse to a single curve if the fit is perfect. We checked that fitting $\rho$ for $J'=1$~mK in the full 3D model reproduces the 2D $T_\mathrm{BKT}$ to within error bars, when fitted at $T>T_\mathrm{LRO}$, where the interlayer coupling becomes insignificant such that the 2D scaling ansatz (\ref{eq:QMC_spin_stiffness}) holds. In Fig.~\ref{fig:QMC_FSS}(a), we show the finite-size collapse of the $\rho$ fit performed at 2 T, for systems with up to one million spins and a temperature grid of $\Delta T=1$~mK. The fit yields $T_\mathrm{BKT}=1.748(15)$~K. 

To determine $T_\mathrm{LRO}$, we consider the scaled in-plane stiffness $L\rho$ for the full 3D model with $J'=1$~mK. At large $L$, this quantity becomes size independent at $T_\mathrm{LRO}$ \cite{sandvik_critical_1998}. Hence, by determining the crossings $T^*$ between $L\rho$ curves with two different sizes $L$, and extrapolating this crossing temperature to $L\rightarrow\infty$, we obtain $T_\mathrm{LRO}$ \cite{luitz_shannon_2014}. 
In Fig.~\ref{fig:QMC_FSS}(b), we show the scaling analysis performed for $L\rho$ at 2 T, where the inset shows the $L\rightarrow\infty$ scaling of the crossing temperature $T^*$. Here, we used a second-order polynomial, which yields $T_\mathrm{LRO}=1.959(2)$~K. Further calculations of the relevant magnetization components and correlation length are presented in Fig.~S3 of the SM \cite{supplemental}.

Employing these procedures at different magnetic fields, we determined $T_\mathrm{BKT} = 1.4877(6)$, $1.7477(15)$, $1.9584(24)$, $1.5323(13)$, and $0.6495(15)$~K, at fields of $0$, $2$, $7$, $12$, and $16$~T, respectively. We also confirmed that $T_\mathrm{BKT}=0$ when both $\Delta=0$ and $H=0$, which emphasizes the strong effect on $T_\mathrm{BKT}$ of  the seemingly small $\Delta=0.0185$ for CuPOF. Furthermore, we determined $T_\mathrm{LRO} = 1.7425(19)$, $1.9597(20)$, $2.1768(23)$, $1.7110(22)$, and $0.7376(17)$~K. At all fields, our calculations yield $T_\mathrm{LRO} > T_\mathrm{BKT}$, thus supporting the experimental phenomenology, as can be seen in Fig.~\ref{fig:phase_diagram}.  We also determined the saturation field to be 17.5~T, in excellent agreement with the experimental value. As in the experiment, the strong dependence of the numerically determined $T_\mathrm{LRO}$ on the field strength reflects the effect of the field-induced anisotropy. The quantitative differences to the experimental transition temperatures at elevated fields might be resolved by extending the complexity of the modelling.
In Fig. S4 of the SM, we obtain a simple estimate of an effective exchange anisotropy $\Delta(H)$ at $H\leq6$~T and compare it to the low-field results \cite{supplemental}.

Our findings suggest the following scenario for the temperature evolution of spin correlations in CuPOF in applied magnetic fields.
Decreasing the temperatures from the paramagnetic high-temperature limit, isotropic Heisenberg-type spin correlations develop which cross over to an anisotropic $XY$-type close to $T_\mathrm{co}$.
With further decreasing temperature, the correlation length $\xi$ grows exponentially due to the vortex physics described by BKT theory.
For $T \agt T_\mathrm{BKT}$, a rather low density of these topological excitations is expected \cite{yosida_theory_1996}.
The exponential increase of $\xi$ yields a rapid strengthening of the antiferromagnetic correlations in the $XY$ regime and, therefore, the staggered magnetization appears effectively non-zero even above $T_\mathrm{LRO}$ (see SM) \cite{supplemental}.
With further increase of $\xi$ upon lowering the temperature further, the magnetic correlations, due to the influence of the small but nonzero interlayer interaction $J^\prime$ on the regions with large in-plane correlation lengths, can no longer be treated as 2D, and a transition to long-range order occurs at $T_\mathrm{LRO}$.
As a consequence of the field-induced BKT-type spin correlations, a concomitant nonmonotonic behavior of the transition temperature $T_\mathrm{LRO}$ is observed experimentally and confirmed by our QMC simulations. 

In conclusion, we found that CuPOF is an experimentally accessible realization of a quasi-2D spin-$1/2$ Heisenberg square-lattice system  with field-tunable magnetic correlations, ranging from almost isotropic Heisenberg to highly-anisotropic $XY$ type.
The phenomenology in CuPOF is driven by field-induced Berezinskii-Kosterlitz-Thouless physics under the influence of extremely small interplane interactions, thus providing an attractive opportunity for systematic investigations of the BKT-type topological excitations and calling for further experimental studies by inelastic scattering techniques.

\section*{Acknowledgments}
We acknowledge the support of HLD at HZDR, member of the European Magnetic Field Laboratory (EMFL).
This work was supported by the Deutsche Forschungsgemeinschaft (DFG) through SFB 1143 (project ID 247310070), the W\"{u}rzburg-Dresden Cluster of Excellence on Complexity and Topology in Quantum Matter -- $ct.qmat$ (EXC 2147, project ID 39085490) and the  cluster of excellence ML4Q (EXC2004, project ID 390534769). Our QMC code is based on the ALPS libraries \cite{alps}. Part of this work was carried out at the STFC ISIS facility and we are grateful for provision of beamtime.
This work is supported by EPSRC (UK). For the purpose of open access, the authors have applied a Creative Commons Attribution (CC BY) licence to any Author Accepted Manuscript version arising. Data presented in this paper resulting from the UK effort is available from \cite{uSRUK}.

\section*{Author contributions}
D.~Opherden and M.~S.~J.~Tepaske contributed equally to this work.


\begin{thebibliography}{99}
\bibitem{footPM}
We note that a constant was added in the fits to account for the paramagnetic contribution to $1/T_{1}$.
\end{thebibliography}
\end{document}


\title{Supplemental Material: Field-tunable Berezinskii-Kosterlitz-Thouless correlations in a 
Heisenberg magnet}

\author{D.~Opherden}
\affiliation{Hochfeld-Magnetlabor Dresden (HLD-EMFL) and W\"{u}rzburg-Dresden Cluster of Excellence ct.qmat, Helmholtz-Zentrum Dresden-Rossendorf, 01328 Dresden, Germany}

\author{M.~S.~J.~Tepaske}
\affiliation{Physikalisches Institut, Universit\"{a}t Bonn, Nussallee 12, 53115 Bonn, Germany}
\affiliation{Max Planck Institute for the Physics of Complex Systems, 01187 Dresden, Germany}

\author{F.~B\"{a}rtl}
\affiliation{Hochfeld-Magnetlabor Dresden (HLD-EMFL) and W\"{u}rzburg-Dresden Cluster of Excellence ct.qmat, Helmholtz-Zentrum Dresden-Rossendorf, 01328 Dresden, Germany}
\affiliation{Institut f\"{u}r Festk\"{o}rper- und Materialphysik, TU Dresden, 01062 Dresden, Germany}

\author{M.~Weber}
\affiliation{Max Planck Institute for the Physics of Complex Systems, 01187 Dresden, Germany}

\author{M.~M.~Turnbull}
\affiliation{Carlson School of Chemistry, Clark University, Worcester, MA 01610, USA}

\author{T.~Lancaster}
\affiliation{Durham University, Centre for Materials Physics, South Road, Durham DH1 3LE, UK}

\author{S.~J.~Blundell}
\affiliation{Clarendon Laboratory, Department of Physics, University of Oxford, Park Road, Oxford OX1 3PU, UK}

\author{M.~Baenitz}
\affiliation{Max Planck Institute for Chemical Physics of Solids, 01187 Dresden, Germany}

\author{J.~Wosnitza}
\affiliation{Hochfeld-Magnetlabor Dresden (HLD-EMFL) and W\"{u}rzburg-Dresden Cluster of Excellence ct.qmat, Helmholtz-Zentrum Dresden-Rossendorf, 01328 Dresden, Germany}
\affiliation{Institut f\"{u}r Festk\"{o}rper- und Materialphysik, TU Dresden, 01062 Dresden, Germany}

\author{C.~P.~Landee}
\affiliation{Department of Physics, Clark University, Worcester, MA 01610, USA}

\author{R.~Moessner}
\affiliation{Max Planck Institute for the Physics of Complex Systems, 01187 Dresden, Germany}

\author{D.~J.~Luitz}
\affiliation{Physikalisches Institut, Universit\"{a}t Bonn, Nussallee 12, 53115 Bonn, Germany}
\affiliation{Max Planck Institute for the Physics of Complex Systems, 01187 Dresden, Germany}

\author{H.~K\"{u}hne}
\email[Corresponding author. E-mail: ]{h.kuehne@hzdr.de}
\affiliation{Hochfeld-Magnetlabor Dresden (HLD-EMFL) and W\"{u}rzburg-Dresden Cluster of Excellence ct.qmat, Helmholtz-Zentrum Dresden-Rossendorf, 01328 Dresden, Germany}

\date{\today}

\maketitle

\subsection{Staggered spin correlations probed by NMR and $\mu^{+}$SR}

\begin{figure}[!ht]
	\includegraphics[width=0.95\linewidth]{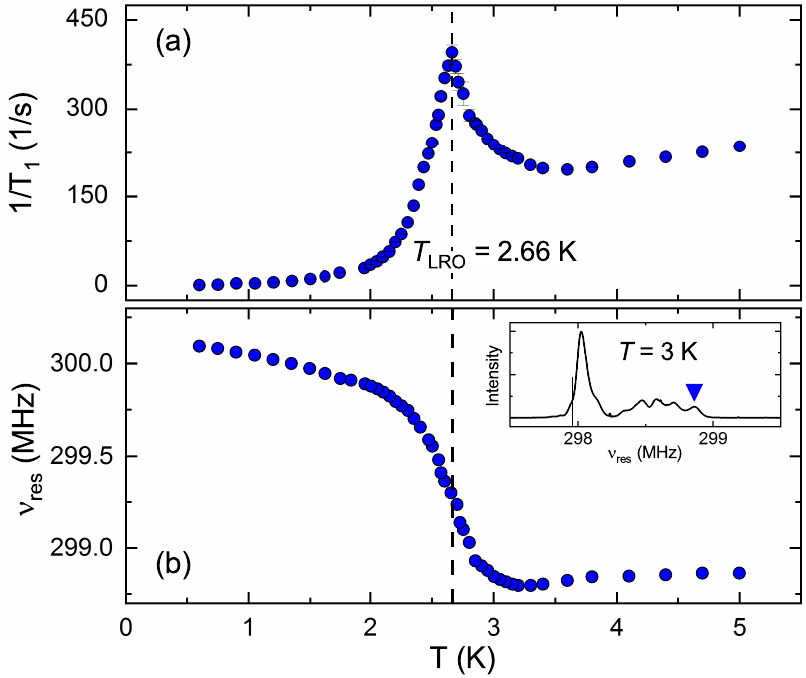}
	\caption{Temperature-dependent $^1$H-NMR (a) spin-lattice relaxation rate and (b) resonance frequency of CuPOF at 7~T. The vertical dashed line indicates the transition temperature $T_\mathrm{LRO}=2.66$~K. A representative $^1$H-NMR spectrum at 3~K is shown in the inset of (b). The blue triangle marks the spectral line under investigation. The solid vertical line indicates the Larmor frequency $\nu_\mathrm{L}= 297.96$~MHz, given by the external field.}
	\label{fig:SM_1H_1overT1_and_res_frequency}
\end{figure}

\begin{figure}[!ht]
	\includegraphics[width=0.90\linewidth]{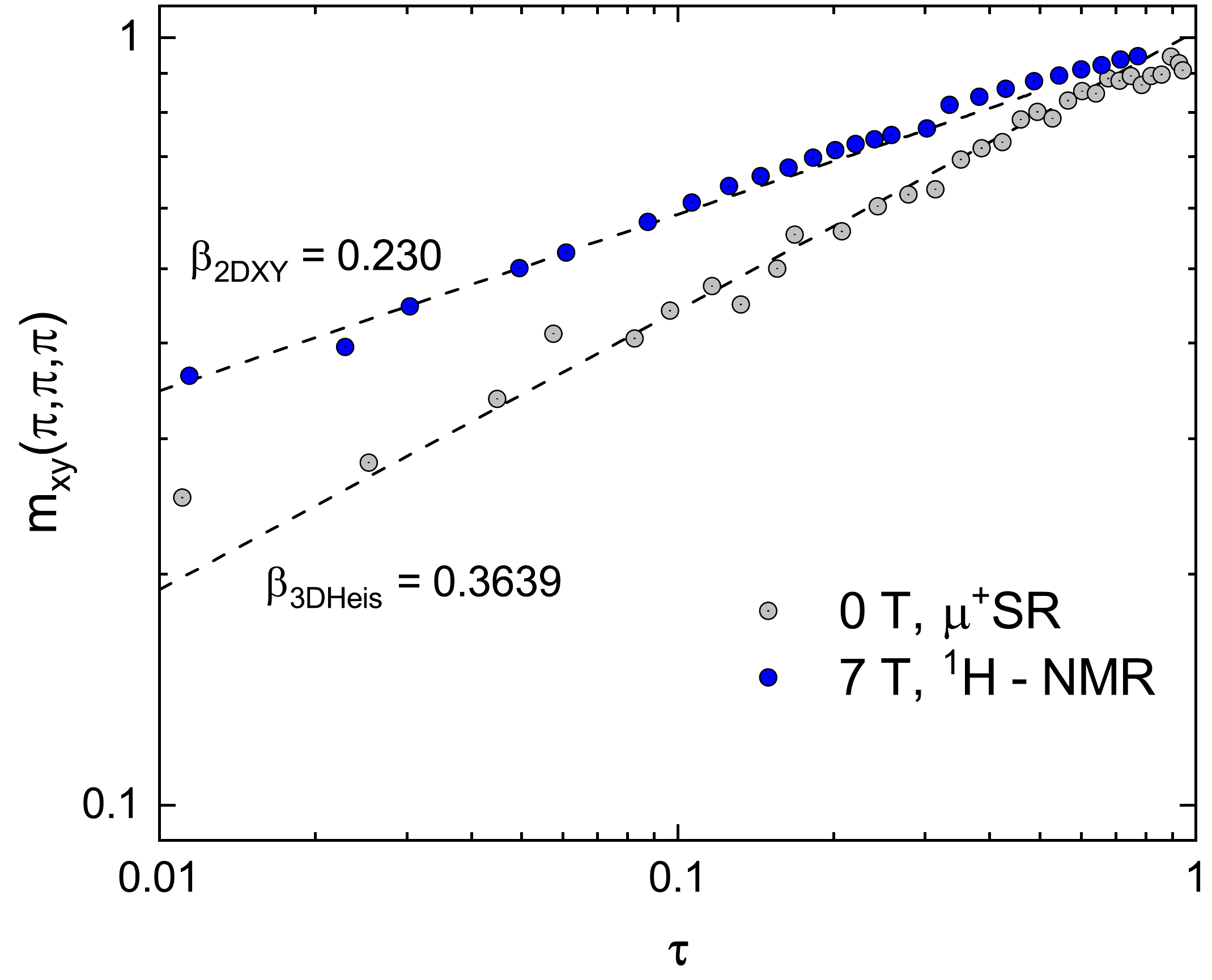}
	\caption{Log-log plot of the normalized $^{1}$H-NMR and $\mu^{+}$SR frequencies ($\mu^{+}$SR data are from Ref.~\cite{opherden_extremely_2020}), probing the staggered magnetization $m_{xy}(\pi, \pi, \pi)$, and plotted as a function of the reduced temperature $\tau = 1 - T/ T_{c}$, with $T_{c}=T_\mathrm{LRO}$. The dashed lines are plots of $m_{xy} \propto \tau^{\beta}$, where  $\beta$ denotes an effective critical exponent.}
	\label{fig:order_parameter}
\end{figure}

In order to investigate the effect of the field-tunable $XY$ anisotropy on the quasi-static spin correlations, we probed the evolution of the staggered magnetization as an effective order parameter.
As reported previously, the $^1$H-NMR spectra of CuPOF yield a distinct line splitting at low temperatures, which provides a direct probe of the local staggered magnetization \cite{opherden_extremely_2020}.
More precisely, the NMR spectrum represents a histogram of the quasi-static fields, probed at the positions of the resonantly excited nuclear moments, on the time scale of the measuring process (a few ten $\mu$s here).
The temperature-dependent resonance frequency $\nu_\mathrm{res}$ (determined as the first spectral moment) of a line at the high-frequency end of the $^1$H-NMR spectrum, recorded at 7~T, is presented in Fig.~\ref{fig:SM_1H_1overT1_and_res_frequency}(b).
The corresponding $^1$H nuclear spin-lattice relaxation rate is presented in Fig.~\ref{fig:SM_1H_1overT1_and_res_frequency}(a), and yields $T_\mathrm{LRO}=2.66$~K, identical to $T_\mathrm{LRO}$ as determined from the $^{31}$P spin-lattice relaxation rate, see Fig. 2(b) in the main text.
A deviation of $\nu_\mathrm{res}$ from an almost constant value at high temperatures occurs at about 3~K $\simeq T_\mathrm{co} > T_\mathrm{LRO}$.

\begin{figure*}[!htb]
	\includegraphics[width=0.45\linewidth]{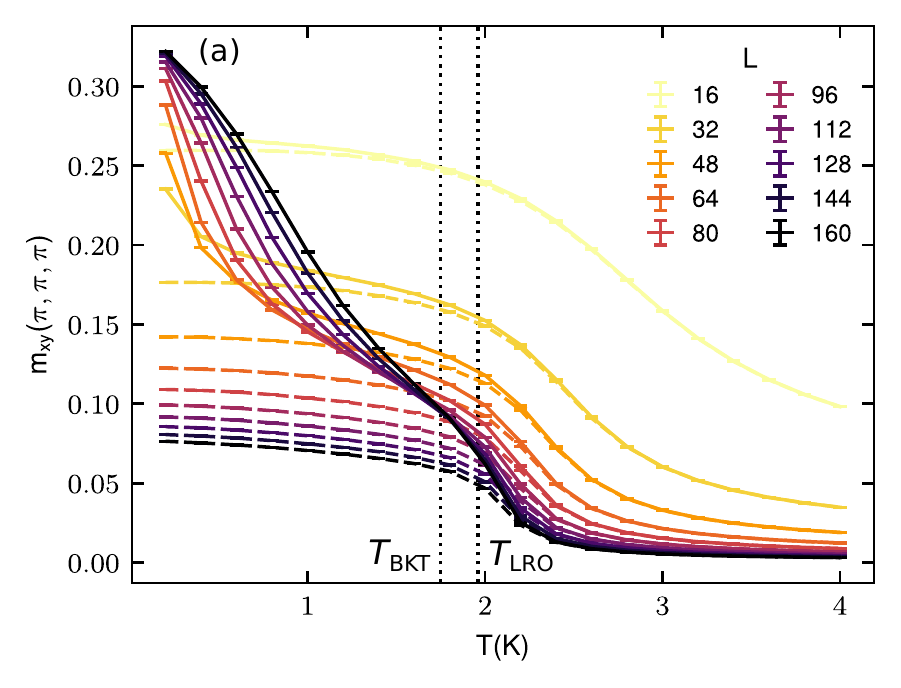}  \vspace{.3cm} \hspace{.3cm}
	\includegraphics[width=0.45\linewidth]{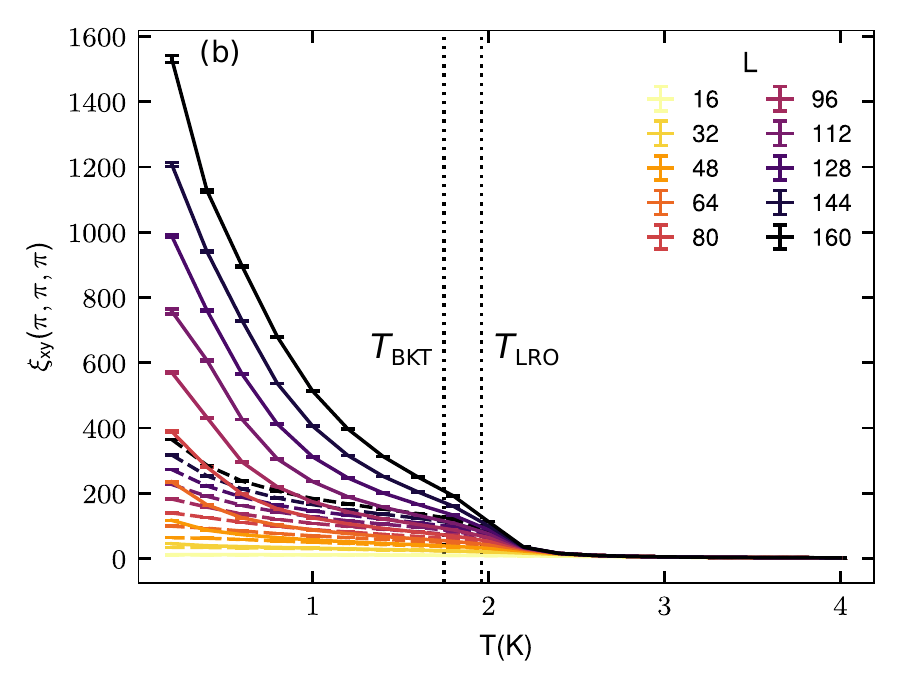} \vspace{.3cm}
	\includegraphics[width=0.45\linewidth]{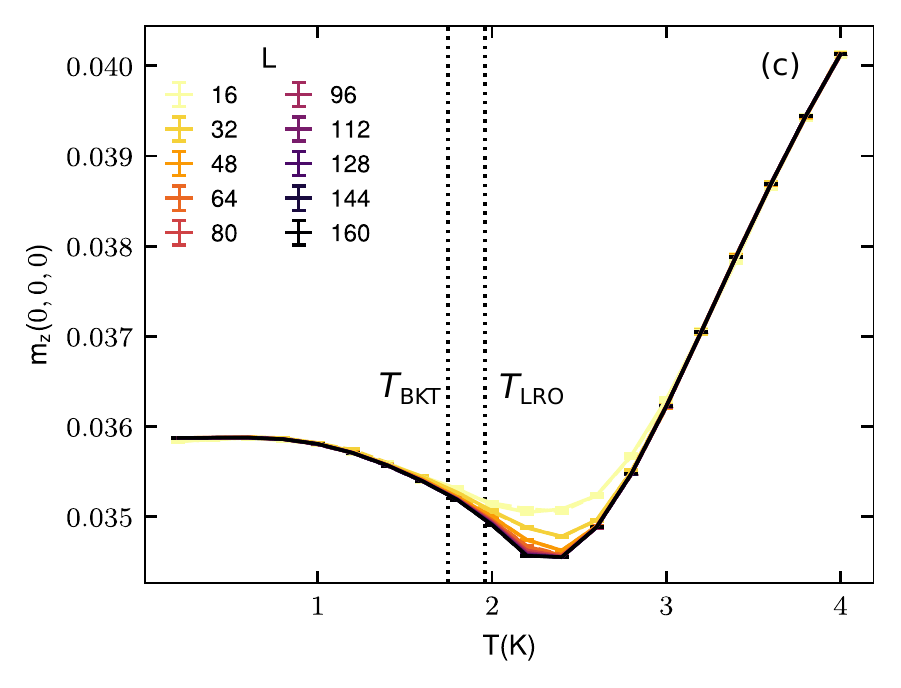} \hspace{.3cm}
	\includegraphics[width=0.45\linewidth]{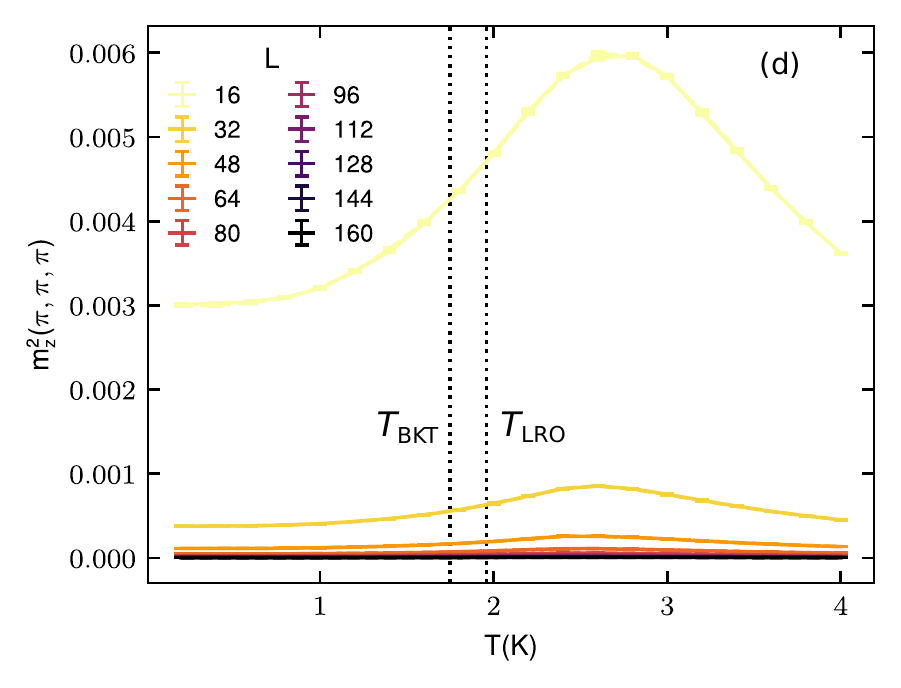} 
	\caption{(a) The staggered in-plane magnetization $m_{xy}(\pi,\pi,\pi)$, (b) the corresponding correlation length $\xi_{xy}(\pi,\pi,\pi)$, (c) the uniform out-of-plane magnetization $m_z(0,0,0)$, and (d) the squared staggered out-of-plane magnetization $m^2_z(\pi,\pi,\pi)$, all determined from the structure factor. The lattices have sizes $L\times L\times L/8$ and we simulated the Hamiltonian (1) in the main text with intralayer coupling $J/k_\mathrm{B} = 6.8$~K, interlayer coupling $J^\prime / k_\mathrm{B} = 1$~mK, intrinsic exchange anisotropy $\Delta = 0.0185$, and magnetic field strength $\mu_0 H=2$~T. The solid lines are for the $J^\prime=1$~mK model and the dashed lines for $J^\prime=0$. The vertical dashed lines denote the critical temperatures $T_\mathrm{BKT}$ and $T_\mathrm{LRO}$ as determined in Fig. 3 in the main text.}
	\label{fig:QMC_mags}
\end{figure*}

Whereas the staggered magnetization already becomes non-zero in the $XY$ regime, 
we define the maximum temperature of the $^{1}$H spin-lattice relaxation rate as critical temperature, see Fig.~\ref{fig:SM_1H_1overT1_and_res_frequency}(a), as supported by the following reasoning.
In a layered anisotropic magnetic lattice, the correlation length significantly increases with decreasing temperature, following an exponential growth described as $\xi_\mathrm{2DXY} \propto \exp \big( 0.5 \pi/\sqrt{ T/T_\mathrm{BKT}-1} \big)$ \cite{ding_could_1992, kosterlitz_critical_1974}.
In the presence of a finite interlayer coupling $J^\prime$, the transition to long-range order is expected at $\xi^2 J^\prime/J \simeq 1$ \cite{als-nielsen_neutron_1993}.
According to our QMC simulations, at $T_\mathrm{LRO}$, the in-plane correlation length is of the order of 100 lattice spacings, as can be seen for $\mu_0 H=2$~T in Fig.~\ref{fig:QMC_mags}(b).
With $J^\prime/J \simeq 1.4 \times 10^{-4}$ for CuPOF \cite{opherden_extremely_2020}, the condition $\xi^2 J^\prime/J \simeq 1$ is satisfied at $T_\mathrm{LRO}$.

Closely below $T_\mathrm{c}$, the staggered magnetization $m_{xy}(\pi, \pi, \pi)$ scales with the reduced temperature $\tau = (1-T/T_\mathrm{c})$ as $m_{xy} \propto \tau^{\beta}$, where $\beta$ may be interpreted as an effective critical exponent. Here, we study to what extent the expected universality classes of an (an)isotropic spin system in (non-)zero field are reflected in the experimental data, but note that there may in practise be a broad crossover region ``interpolating'' between the two (Heisenberg and XY) cases \cite{isakov_2003}.

\begin{figure*}[!htb]
	\includegraphics[width=0.9\linewidth]{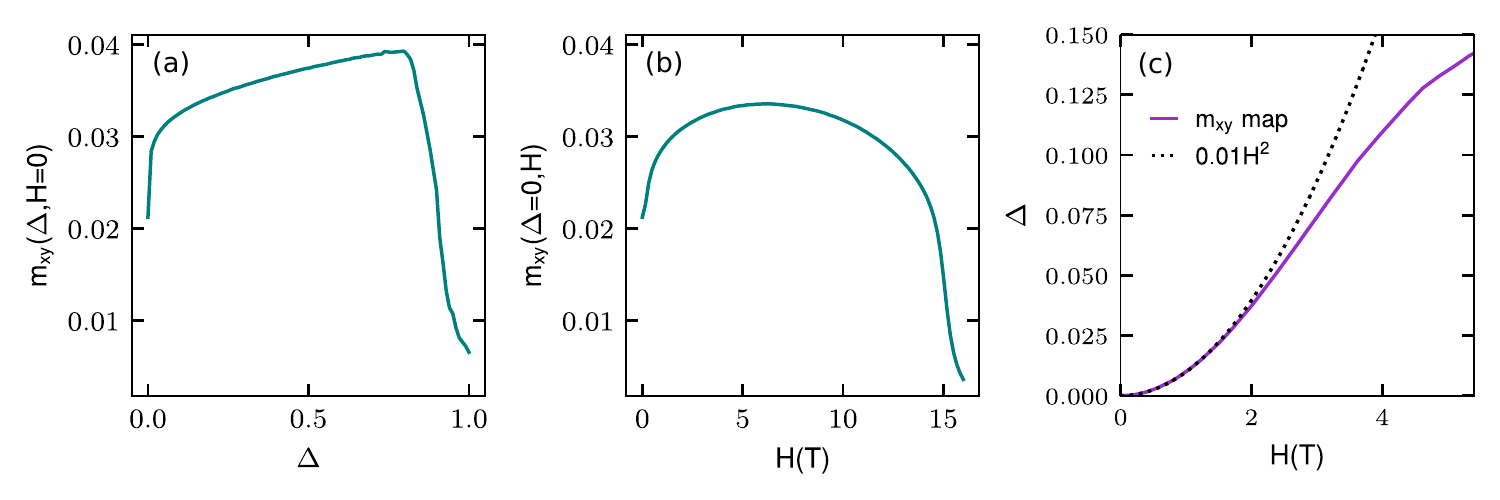}
	\caption{(a) The staggered in-plane magnetizations $m_{xy}(\Delta,H=0)$ and (b) $m_{xy}(\Delta=0,H)$ for an $L\times L = 100\times100$ Heisenberg system at $T=1.2$~K. In (c), we show the numerically determined field-induced exchange anisotropy $\Delta(H)$, that was determined for $\mu_0 H\leq6$ T by solving $m_{xy}(\Delta,H=0)=m_{xy}(\Delta=0,H)$. The black dotted line shows the quadratic dependence with $\Delta=0.01 (\mu_0 H)^2$.}
	\label{fig:mxy_DH}
\end{figure*}

Employing $T_c = T_\mathrm{LRO} = 2.66$~K at 7~T, we plot the normalized $^{1}$H resonance frequency as a function of the reduced temperature in a log-log plot, see Fig.~\ref{fig:order_parameter}.
We find a good agreement when comparing with the critical exponent $\beta_\mathrm{2DXY} = 3\pi^2/128 \simeq 0.23$ of a finite-size 2D $XY$ model \cite{bramwell_magnetization_1993,bramwell_magnetization_1994}. Similar observations were made for other materials that realize a planar $XY$ lattice \cite{taroni_universal_2008, klyushina_signatures_2021}.
The same analysis was applied to the $\mu^{+}$SR frequency at zero field \cite{opherden_extremely_2020}, using $T_\mathrm{c} = 1.38(2)$~K, giving a good agreement when comparing to the critical exponent $\beta_{\text{3DHeis}} = 0.3639(35)$ \cite{troyer_critical_1997} of the 3D Heisenberg model, and, similarly well, to the critical exponent $\beta_\mathrm{3DXY} = 0.33$ \cite{als-nielsen_neutron_1993} of the 3D $XY$ model. 
These observations further support the scenario of the enhanced anisotropy of intralayer spin correlations at elevated fields.

\subsection{Staggered spin correlations calculated with QMC}

To infer the pattern of the long-range order that is associated to $T_\mathrm{LRO}$, which was calculated with QMC by examining the $L\rho$ crossings in the lower panel of Fig. 3 in the main text, we also calculate the in-plane structure factor $S_{xy}$ and the corresponding correlation length $\xi$ \cite{sandvik_computational_2010}, i.e.,
\begin{align}
    S_{xy}(\vec{k}) &= \sum_je^{i\vec{k}\cdot\vec{r}_j}\left(\langle S^x_0S^x_j\rangle+\langle S^y_0S^y_j\rangle\right), \\
    \xi_{xy}(\vec{k}) &= \frac{L}{2\pi}\sqrt{\frac{S_{xy}(\vec{k})}{S_{xy}(\vec{k}+\vec{\text{d}k})} - 1} \, .
\end{align}
Here, we introduced a staggering phase based on the lattice position $\vec{r}_j$ of site $j$ and a staggering vector $\vec{k}$ with nearest-by vector $\vec{k}+\vec{\mathrm{d}k}$. This structure factor can be used to define a magnetization via $m^2_{xy}=S_{xy}(\vec{k})/N$, where $N$ denotes the amount of spins. In Fig.~\ref{fig:QMC_mags}(a), we show the staggered in-plane magnetization $m_{xy}(\pi,\pi,\pi)$ at 2 T as a function of temperature for various values of $L$, and in Fig.~\ref{fig:QMC_mags}(b) we show the corresponding correlation length $\xi_{xy}(\pi,\pi,\pi)$. The $J'=1$~mK results are shown as solid lines and the $J'=0$ results as dashed lines. We see an onset of the magnetization and in-plane spin-correlations at $T_\mathrm{LRO}$ that does not scale to zero with system size when $J'$ is nonzero. 

In Fig.~\ref{fig:QMC_mags}(c), we show the uniform out-of-plane magnetization $m_z(0,0,0)$, which depicts the field dependency of the out-of-plane canting of the in-plane antiferromagnetic order and converges with system size. To verify that the magnetic order is only in-plane staggered, we show the squared staggered out-of-plane magnetization $m_z(\pi,\pi,\pi)$ in Fig. \ref{fig:QMC_mags}(d), which clearly scales to zero for large system sizes.

\subsection{Field-induced exchange anisotropy} \label{anisotropy_numerical}

As a simple estimate of the field-induced exchange anisotropy, we compute the staggered in-plane magnetizations $m_{xy}(\Delta,H=0)$ and $m_{xy}(\Delta=0,H)$ for an $L\times L = 100\times100$ Heisenberg system at $T=1.2$~K, and find $\Delta(H)$ such that $m_{xy}(\Delta,H=0)=m_{xy}(\Delta=0,H)$ is satisfied. This condition should hold if the Hamiltonians $\mathcal{H}(\Delta,H=0)$ and $\mathcal{H}(\Delta=0,H)$ can be mapped onto each other, thereby giving an estimate of the field-induced exchange anisotropy $\Delta(H)$. In Figs.~\ref{fig:mxy_DH}(a) and \ref{fig:mxy_DH}(b), we show $m_{xy}(\Delta,H=0)$ and $m_{xy}(\Delta=0,H)$, and in Fig.~\ref{fig:mxy_DH}(c) we show the numerically determined field-induced exchange anisotropy $\Delta(H)$ that was found at $H\leq6$~T. To compare with the perturbative quadratic field dependence estimate from \cite{sengupta_nonmonotonic_2009}, we also plot the quadratic curve $\Delta=0.01 (\mu_0 H)^2$, showing excellent agreement at small $H$.